REPORT

# AI could create a perfect storm of climate misinformation

Victor Galaz, Hannah Metzler, Stefan Daume,
Andreas Olsson, Björn Lindström and Arvid Marklund

Stockholm Resilience Centre | Stockholm University

Beijer Institute OF ECOLOGICAL ECONOMICS | KUNGL. VETENSKAPS-AKADEMIEN THE ROYAL SWEDISH ACADEMY OF SCIENCES

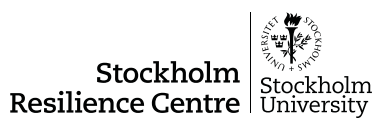
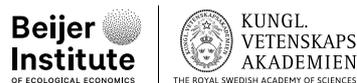
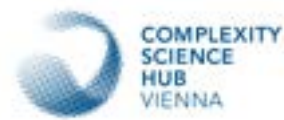
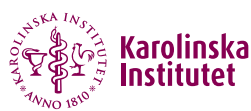

With support from:

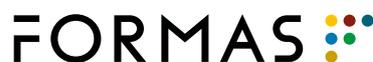
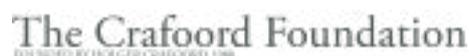

# Content

# AI could create a perfect storm of climate misinformation

**We are in the midst of a transformation of the digital news ecosystem. The expansion of online social networks, the influence of recommender systems, increased automation, and new generative artificial intelligence (AI) tools are rapidly changing the speed and the way misinformation about climate change and sustainability issues moves around the world. Policymakers, researchers and the public need to combine forces to address the dangerous combination of opaque social media algorithms, polarizing social bots, and a new generation of AI-generated content.**

**When OpenAI released their version** of the artificial intelligence ChatGPT to the public in November 2022, the playing ground for false news and misinformation online changed completely. Suddenly, anyone could with relatively little effort produce seemingly human-written text and highly realistic images – regardless of their truthfulness.

Misinformation and disinformation campaigns are not new, and topics such as climate change have been in the line of fire more than once. But over time, the digital aspects of these phenomena have become increasingly important for those interested in understanding and tackling the harmful effects of mis- and disinformation.

Information and communications technologies have made the world increasingly connected, allowing information to move almost effortlessly around the world in the blink of an eye. What we see on our screens everyday is increasingly determined by recommender systems, systems designed to maximize engagement, at times favoring whatever creates engagement over truth. Sometimes, this information movement is affected by social bots. Often, the diffusion of information is shaped by how people engage with it by sharing, commenting, remaking it and moving it across digital media platforms and between social and conventional media. This, in combination with the capacities of generative AI to create synthetic content, may very well result in a perfect storm of misinformation. The time to act on these risks is now.

The following synthesis explores the increasing influence of digital media, AI and algorithmic systems on the creation, diffusion and amplification of misinformation about climate and environmental sustainability issues. We combine insights from various research strands ranging from computational social sciences, to sustainability sciences and neurosciences, to explore what we know about how such information is amplified, and to what extent recent advances in AI pose new challenges for our ability to act collectively on our planetary crisis.

This synthesis brief has been put together as an independent contribution to the Nobel Prize Summit 2023, "Truth, Trust and Hope", Washington D.C., 24-26 of May 2023.

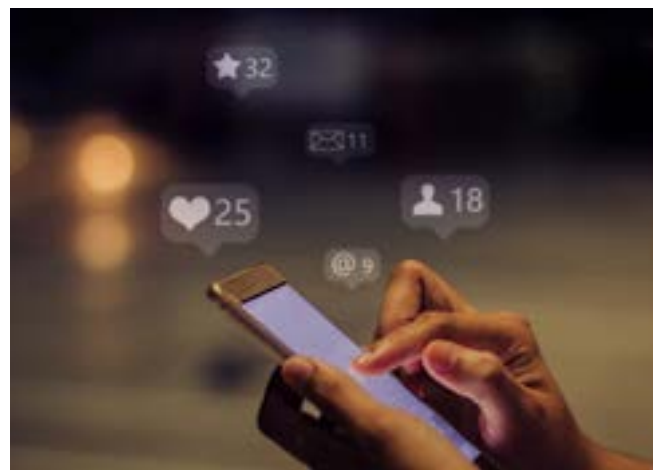

Image: Canva

## Box 1. Key terms

**Artificial intelligence** – we use the terms "artificial intelligence" and "AI" to refer to technologies that employ machine learning including deep learning methods.

**Algorithmic systems** – refer to assemblages of multiple algorithms (i.e., a finite sequence of well-defined, computer-implementable instructions, typically to solve a class of problems or to perform a computation), databases, and interfaces that interact, as well as the social practices that people develop to interact with them (Wagner et al., 2021; Ada Lovelace Institute, 2021).

**Digitalization** – is here used to describe the adoption and use of digital technologies, including societal, organizational, and individual impact (Legner et al., 2017). Digital technologies for e.g., food production ('precision farming'), electricity ('smart grids'), and housing ('smart homes'), do not necessarily use AI for their operations.

**Generative AI** – refers to AI based on advanced neural networks with the ability to produce highly realistic synthetic text, images, video and audio (including fictional stories, poems, and programming code) with little, to no human intervention. The most known example is the Generative Pre-trained Transformer 3 (GPT-3, GPT-4) that underpins OpenAI's GPTChat (OpenAI, 2023). Note however that GPT is only one of many existing generative AI models (e.g., Gibney, 2022).

**Recommender systems** – refer to techniques that offer suggestions for information (e.g., other users, text, articles, videos or social media posts) that are likely to be of interest to a particular user. Recommender systems are widely used in social media platforms, but also other digital services like music streaming, video services, and for online shopping (Ricci et al., 2015). Such systems can deploy AI-analysis to guide their recommendations (Engström and Strimling, 2021).

**Social bots** – are automated social media accounts with the ability to build social communication networks and create online content. Such bots can be used for beneficial or malicious purposes ranging from news aggregation, user assistance or entertainment, to spam and sophisticated influence campaigns (Stieglitz et al., 2017).

# What is climate mis- and disinformation – and why should we care?

By Victor Galaz

**Our living planet is facing a multitude of interacting crises. Climate change, the loss of biodiversity, inequality, pandemics, and war and conflict are just a few of them. In these times of turbulence, it is vital that decisions about our common future are well-grounded in science. False information and intentional attempts to manipulate public opinion pose serious risks to our joint capacities to create a safe and just future for all.**

**Mis- and disinformation play a crucial** role in forming public opinion and thus influencing the actions taken at all levels of society, from individuals to central policy-making. Misinformation about covid-19 vaccines for example, has notable impacts on whether people are willing to take the vaccine (Loomba et al., 2021). A growing body of work also explores to what extent misinformation has affected political elections around the world (for example, Howard, 2020; Machado et al., 2019). Scientists engaged in science communication also face an increased number of attacks in social media as they struggle to inform the public to mitigate the spread of fake news and misinformation (Nogrady, 2021).

It is becoming increasingly clear that misinformation is taking its toll on public opinion on climate and sustainability issues as well. A recent study shows that U.S. citizens, partly due to biased and incorrect media reporting, systematically underestimate public support for ambitious climate policies, thus falsely assuming that a vocal minority who dismiss climate change are representative of the broader public opinion (Sparkman et al., 2022). A cross-national study conducted by the Climate Action Against Disinformation and Conscious Advertising Network (2022) shows a considerable spread of serious misconceptions about climate change in the surveyed countries. 33% of the surveyed population in the US and Australia believe that 'the climate has always changed, global warming is a natural phenomenon and is not a direct result of human activity.' In Brazil, 30% believe that climate change is not caused mainly by human activity, and 24% believe that the "temperature record is unreliable or rigged". In the United Kingdom, 29% believe that a significant number of scientists disagree on the cause of climate change (Climate Action Against Disinformation and Conscious Advertising Network, 2022). Misinformation and conspiracy theories about wind energy farms are already affecting the expansion of renewable energy negatively, and thus the prospects for achieving a transition to zero-carbon energy sources (Winter et al., 2022).

One key and challenging issue is how to define and identify mis- and disinformation. Here we refer to misinformation as information whose inaccuracy is unintentional. Disinformation on the other hand, refers to information that is deliberately false or misleading (Jack, 2017). While the definitions of mis- and disinformation might seem straightforward, their differences in the real world are not (Jerit and Zhao, 2020; Treen et al., 2020).

Climate mis- and disinformation has a long history, and its underlying motivations and strategies have become increasingly visible in the last years (e.g., Supran et al., 2023; Franta, 2018). Attempts to sow confusion about the science of climate change include lobbying campaigns by vested interests (Brulle, 2018), the funding of climate change denialist think tanks

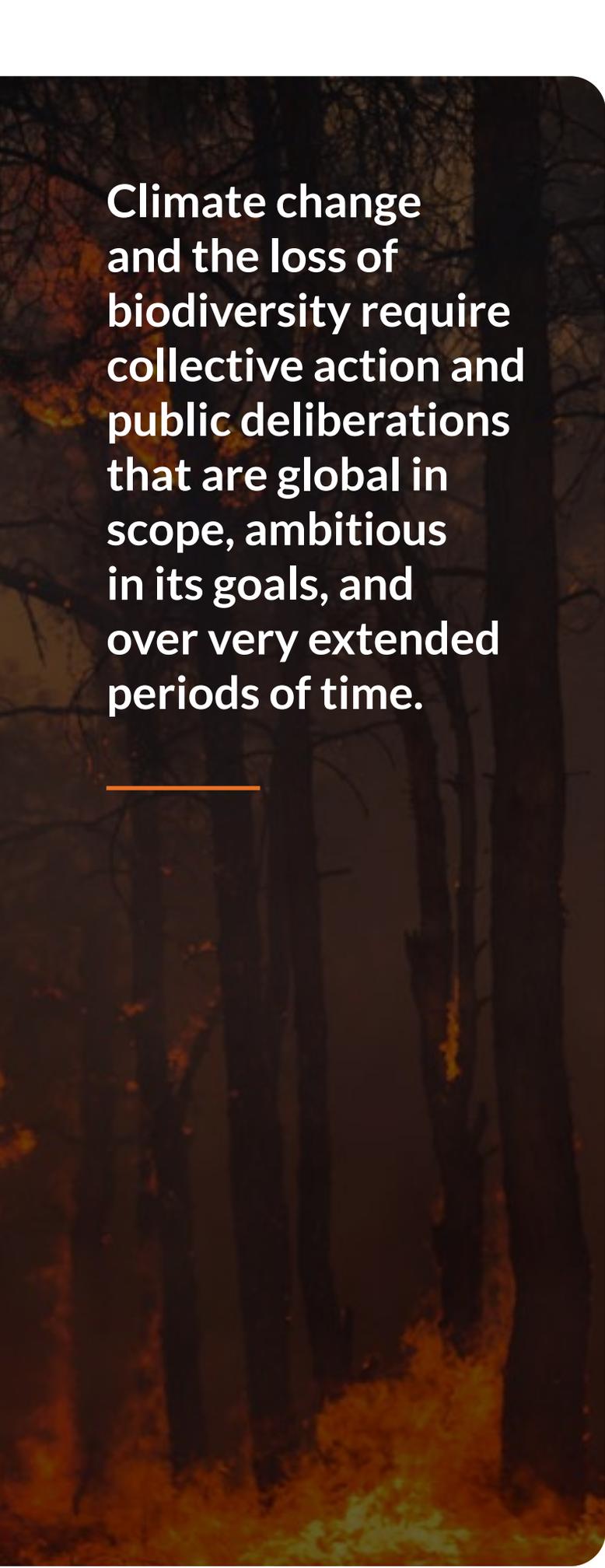

**Climate change and the loss of biodiversity require collective action and public deliberations that are global in scope, ambitious in its goals, and over very extended periods of time.**

Image: Canva

(Farrell, 2019), and corporate climate "skeptic" campaigns in conventional media (Dunlap & McCright, 2011; Supran & Oreskes, 2017).

In parallel to this type of coordinated disinformation activities there are other much more fluid and complex self-organized patterns of information sharing and engagements on digital platforms. In the case of fossil fuel companies' documented attempts to influence public opinion on climate change issues their agency is straightforward (Supran and Oreskes, 2017; Supran et al., 2023). But the type of mis-and disinformation that flows effortlessly across social media platforms like Twitter, TikTok and Youtube, is another type of beast. The former has a clear intent, plan and coordinating agent, while the latter form tends to emerge and evolve as information, algorithmic systems and digital social networks interact in complex ways. These two forms of mis- and disinformation are of course, often combined in reality (Starbird and Wilson, 2019).

Digital media plays a key role in this regard (Pearce et al., 2019; Treen et al., 2020). Five hundred million tweets, 294 billion emails, 4 petabytes of content on Facebook, 65 billion messages on WhatsApp, and 5 billion searches on Google are conducted every day (from Howard, 2020, p. 4). Social media like Facebook, Twitter and Instagram is particularly important in this context. Survey data shows that between 40% and 60% of adults in most developed countries, receive their news from social media (Nic, 2017). The fact that social media platforms are participatory by design allows them to shape individual attitudes, feelings and behaviors (Williams et al., 2015), and facilitates social mobilization and protests (Steinert-Threlkeld et al., 2015).

Climate change and the loss of biodiversity require collective action and public deliberations that are global in scope, ambitious in its goals, and over very extended periods of time (Galaz 2020). Polarization, mistrust in science, and incorrect climate and environmental information amplified through digital media could undermine such much-needed collective action in detrimental ways.

# The neuroscience of false beliefs

By Andreas Olsson and Björn Lindström

---

**What we believe is not only a result of our own reasoning, but also of the beliefs of people around us. Our brains are wired to consume information that is liked by our peers. That way, social reinforcers online – in the form of likes, comments and shares – can build a basis for what individuals believe to be true or false.**

**Why do people adopt beliefs,** such as that the earth is flat, which are clearly incompatible with established truth? Like with other beliefs, it depends on the nature of their prior beliefs, goals, as well as learning and reasoning processes. Our knowledge about the psychological and neural mechanisms that promote false beliefs is still nascent, but existing research in psychology and neuroscience has described how people process information that is unexpected, emotional, and politically divisive – features typical of false information encountered on social media.

First, information that is unexpected and emotionally (especially negatively) loaded, tends to grasp our attention, which depends on visual cortices and attentional networks in parietal and frontal brain regions. In addition, attention also depends on activity in deep brain nuclei that are sensitive for events that can be threatening and important goals of the individual.

Various goals of the individual can be important. For example, if social goals, such as promoting one's own social status or belonging to a particular group, are more important than the goal to seek and represent the truth, then attention is biased towards whatever promotes the more important goals. Research has shown that the orbitofrontal cortex, a brain region located just behind the eyes, is critical for computing the value of different goals. Other prefrontal regions serve to monitor information that deviates from what is in line with one's own moral or political convictions. People who strongly identify with different political ideologies tend to distort their conclusions in line with their political identities when asked to reason about

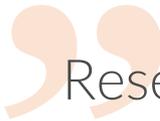

> Research has shown that the orbitofrontal cortex, a brain region located just behind the eyes, is critical for computing the value of different goals.

politically divisive facts. At first glance, one could think that this is due to politically motivated reasoning. However it can also be viewed as rational inferences considering the individual's prior belief system (Botvinik-Netzer et al., 2023). Networks of brain regions linked to memory retrieval and belief updating, are likely to support these processes.

Recent research shows that simple learning mechanisms, known to engage regions in the brain's system that run on dopamin, can explain how humans are motivated to increase the consumption of information that is liked by others – in particular by those belonging to the same group as the individual identifies with. For example, models of instrumental learning explain why people engage with social media (Lindström et al., 2021, Ceylan et al., 2023) and how morally outraged messages become viral (Brady et al., 2021). The social reinforcers that are at play in social media might thus provide a basis for what individuals believe to be true. Indeed, research has shown that social forms of learning can be at least as powerful, and draw on many of the same neural principles as learning based on own, personal experiences (Olsson et al., 2020). The importance of understanding the power of social influences is further underscored by recent findings showing that the more a statement is "liked" by others, the more the statement is subsequently rated as true (Granwald et al., 2023).

In sum, although much is known about the psychological and brain bases of acquiring and updating beliefs in general, much less is known about how we process false beliefs encountered online. Because the lack of a shared knowledge about what is true can fuel polarization, conflicts and other destructive behaviors (Brady et al., 2023), we need to increase efforts to understand the basic psychology and neuroscience of false beliefs.

> Our brains are wired to consume information that is liked by our peers.

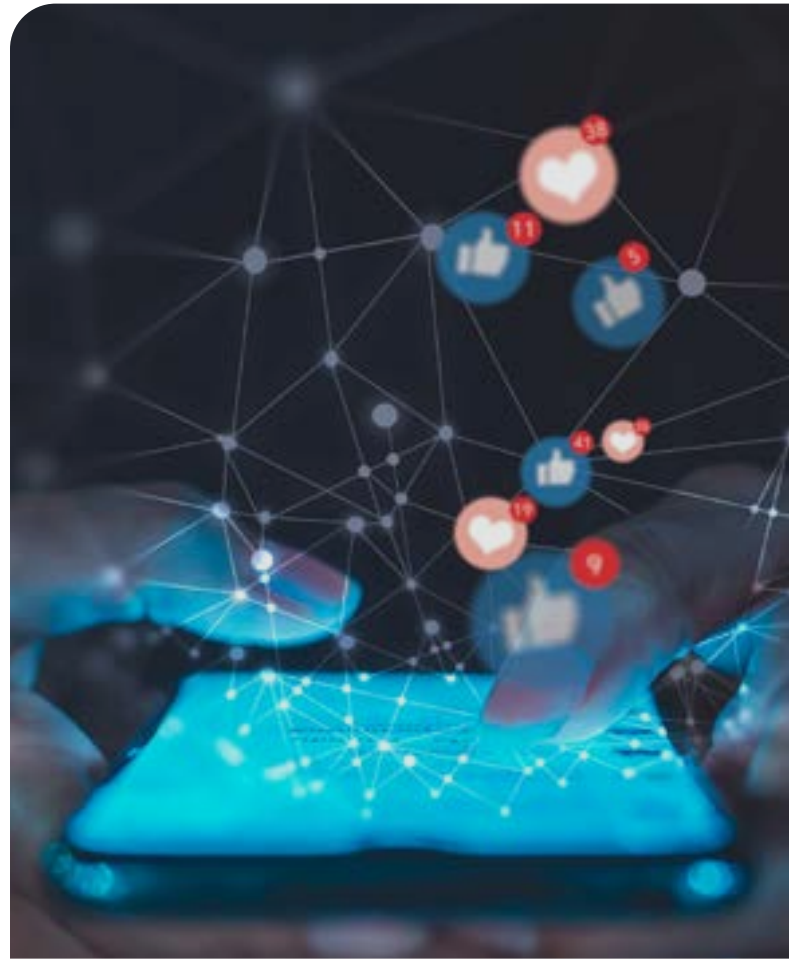

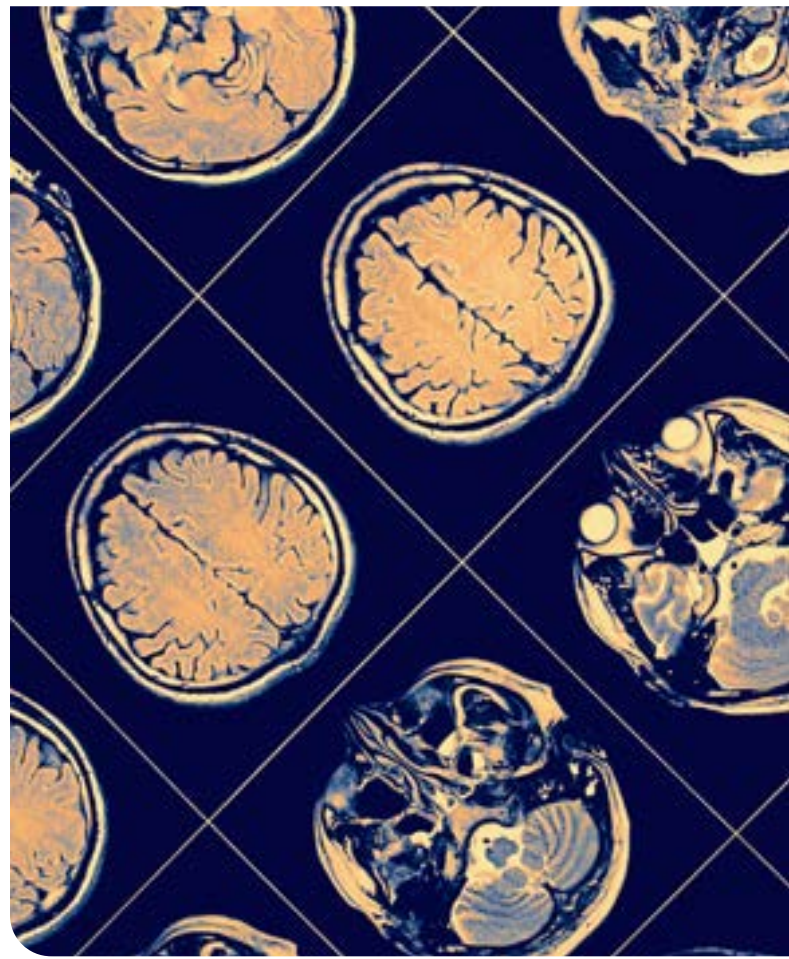

Image: iStockphoto / Canva

# How algorithms diffuse and amplify misinformation

By Victor Galaz and Stefan Daume

**Digital platforms are designed to maximize engagement. Recommender systems play a key role as they shape digital social networks, and the flow of information. Weaknesses in their underlying algorithmic systems are already today being exploited to try to influence public opinion on climate and sustainability issues. Automation through social bots also play a role, although their impacts are contested.**

**In 2010, a team of political scientists** in collaboration with researchers at Facebook conducted a gigantic experiment on the platform. It included "all users of at least 18 years of age in the United States who accessed the Facebook website on 2 November 2010, the day of the US congressional elections". This 61-million-person experiment tested whether a simple tweak in the newsfeed of users – adding a "I Voted" button and information about friends who had voted – could impact voting behavior.

The results were clear: this small design change led to higher real-world voting turnout. Facebook-users who saw a message about one of their close friends on Facebook having voted were 2.08% more likely to vote themselves (Bond et al., 2012).

This controversial experiment illustrates a number of issues that are key if we are to understand how digital platforms interplay with the creation and diffusion of mis- and disinformation, and their impacts.

First, digitalization and the expansion of social media has not only fundamentally expanded the scale, but also transformed the properties of social networks. 3,6 billion people use social media today. Through it, they access a digital media environment where language barriers slowly but surely are eroding as new digital tools make communication across borders increasingly frictionless. These new connections are transforming the structures of social networks in ways that allow for immediate communication across vast geographical space (Bak-Coleman et al., 2021).

Second, the way information moves in these vastly spanning digital social networks is not random. Its movement is fundamentally affected by the way digital media platforms such as recommender systems are designed to maximize engagement and induce other forms of behavioral responses amongst its users (Kramer et al., 2014; Coviello et al., 2014). The diffusion of such information can also be exploited by external attempts to manipulate public opinion online. Three common approaches include (1) specifically targeting the way algorithmic systems operate, (2) carefully crafting viral messages to sow confusion, and (3) through the use of automation tools like 'social bots'.

Social bots mimic human behavior online, and can be used in ways to amplify certain types of information – say, climate denialism –, or operate in ways that widen social divisions online (Gorwa and Guilbeault, 2018; Shao et al., 2018). Our own work (Daume et al., 2023) shows that social bots play an unignorable role in climate change conversation. They amplify information that supports and opposes climate action at the same time, especially information appealing to emotions, such as sympathy or humor.

Recommender systems play a key role for the dynamics of information diffusion and the evolution of digital social networks (Narayanan, 2023), and they have increasingly become infused with deep learning-based AI (Engström and Strimling, 2020). People-recommender systems for example (like "People You May Know" on Facebook or "Who to Follow" on Twitter) shape social network structures, thus influencing the information and the opinions a user is exposed to online (Cinus, 2021). Content-recommender systems (like "Trends for you") can reinforce the human preference for content that aligns with a user's ideology (Bakshy et al., 2015).

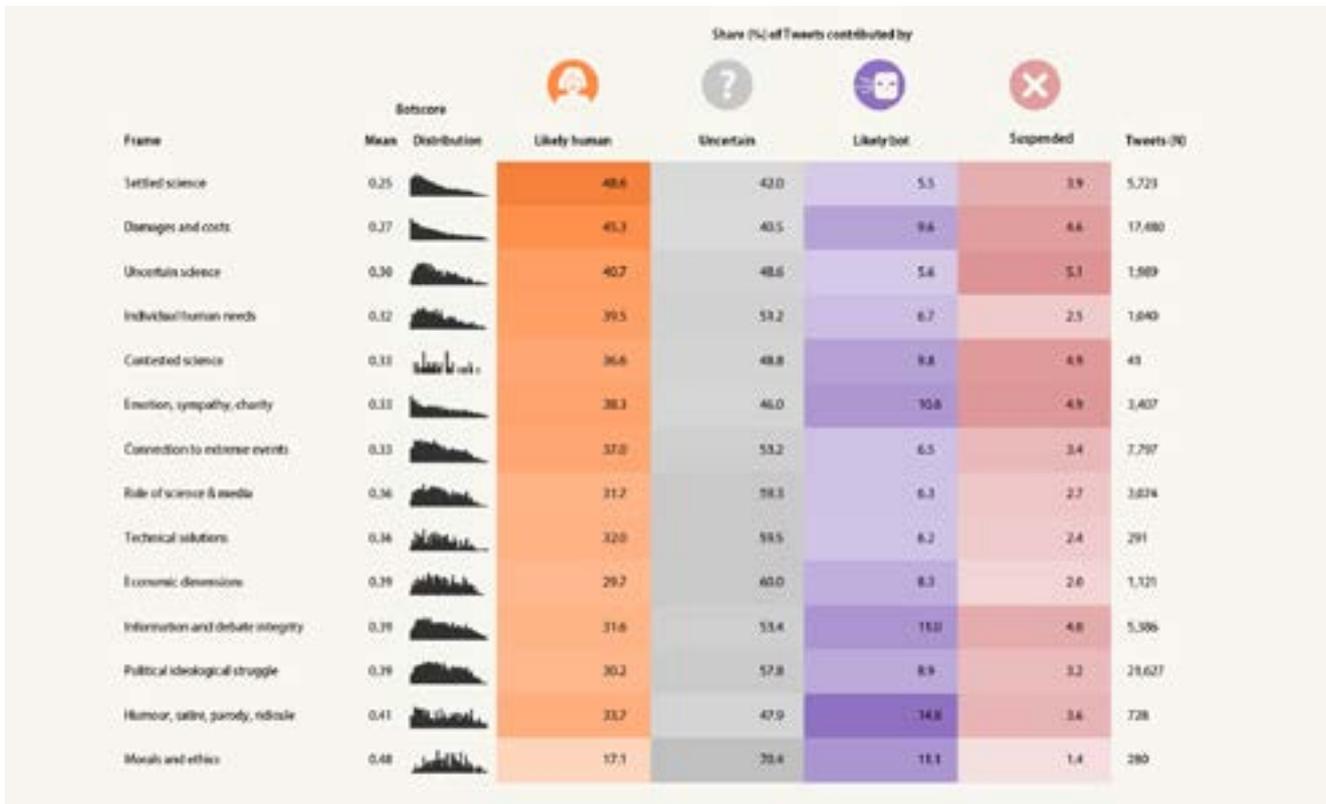

**Figure 1.** The figure illustrates how social bots influenced the diffusion of different climate change frames via Tweets about the 2019/2020 Australia bushfires. The results indicate that different framings of climate change are associated with distinctive automation signatures.

Figure 2 below illustrates four highly simplified models of information propagation of one individual post. The expansion of digital social networks, the influence of recommender systems and social bots change the reach of mis- and disinformation. This diffusion has complex secondary effects on perceptions, on the formation of online communities, on collective action, and on identity formation.

Lastly, the 61-million-user experiment also illustrates another feature of today's digital ecosystem – the capacity to mass produce and test online material to maximize impact. Online material and digital platforms provide the perfect setting to conduct randomized controlled experiments as a means to systematically compare two versions of something – say, a news article or ad – to figure out which performs better, sometimes referred to as A/B testing. In combination with a growing capacity to cheaply and quickly mass produce synthetic text, images and videos using generative AI (see chapter 6), this is pushing us into uncharted and dangerous territory for the future of mis- and disinformation on climate and environmental sustainability issues.

> People-recommender systems for example (like "People You May Know" on Facebook or "Who to Follow" on Twitter) shape social network structures, thus influencing the information and the opinions a user is exposed to online.

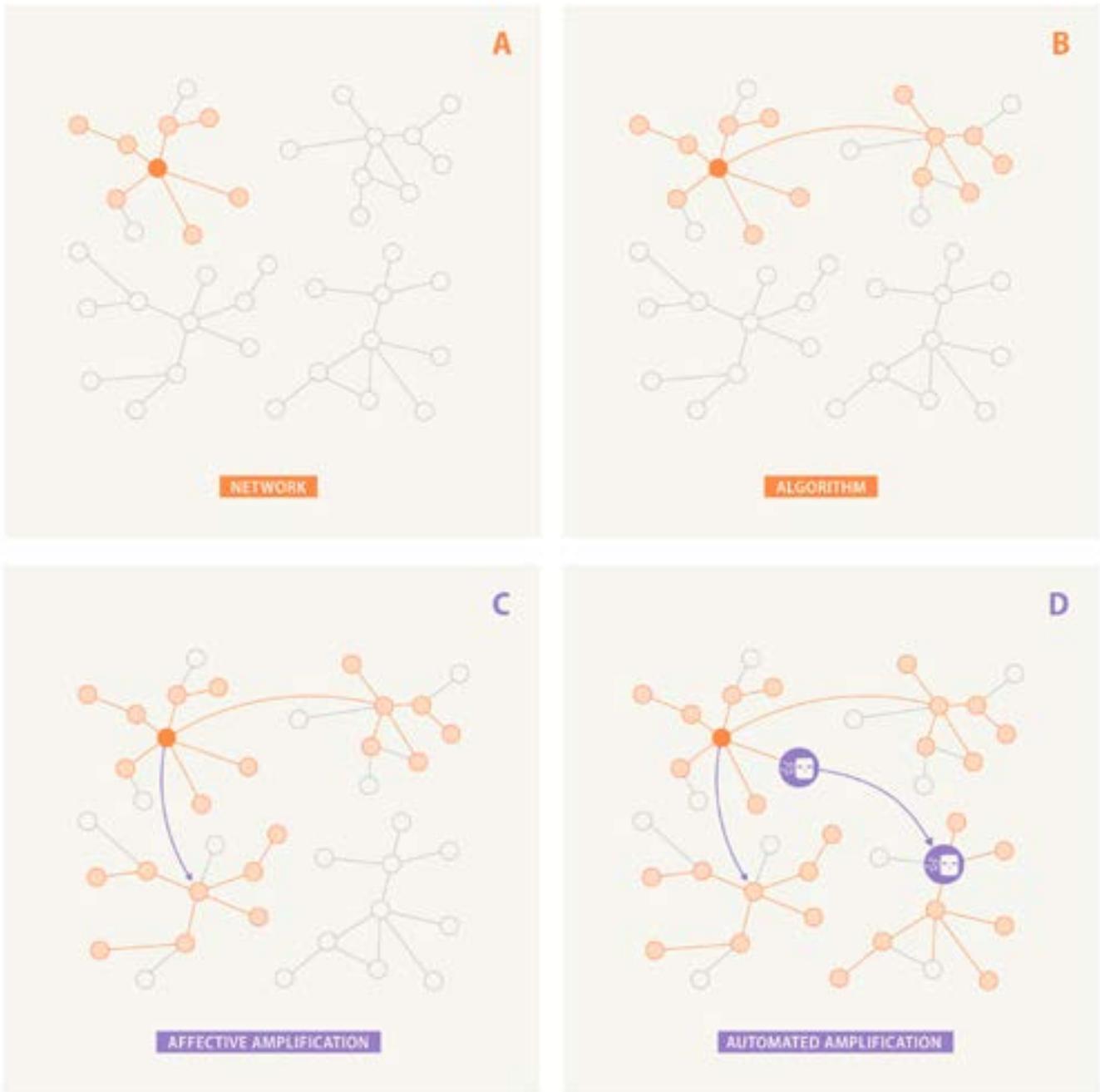

**Figure 2.** The figure illustrates four models of information propagation of one individual post through: a) network, b) algorithm, c) affective, and d) automated amplification. In a) the post cascades through the network as long as other users choose to further propagate it by e.g., sharing or liking. In the algorithmic model b) propagation unfolds as users with similar interests (as determined by recommendation algorithms based on for example past engagement) are more likely to be recommended the post. In (c), users comment and reshare posts much more if they elicit strong emotions, and recommender systems pick up highly engaging posts and amplify them even further. In d) automated accounts ('social bots') purposefully share content that elicits strong emotions to further increase propagation. Figure based on (Narayanan, 2023).

# Emotions and group dynamics around misinformation on social media

By Hannah Metzler

**Nuanced views and clean facts don't generate clicks. Social networks are designed to speak to our emotions, and the more extreme the emotions, the better the content. But that does not mean that good arguments, education, and science communication are futile.**

**Emotions attract our attention,** and provide us with information about actions we should take: When you feel fear, it's best not to ignore the danger, and protect yourself from it. When you are angry, it's probably because someone has treated you or a group you belong to unfairly, and it's time to step up against the injustice.

News agencies, politicians, and creators of fake news know this, and use it to create content that attracts attention and is likely to be shared on digital and social media. Algorithms on social media are optimized to increase engagement with content (Narayanan, 2023; Metzler & Garcia, 2022), and content that provokes strong emotional reactions is a powerful means to do so. Negative moral-emotional messages about groups we do not like seem to particularly increase engagement (Brady et al., 2017, Rathje et al. 2021, Marie et al., 2023). Humans are social animals, and things that make us feel part of a group, that increase our group's status, or decrease the status of an out-group, are highly motivating for us (Robertson et al., 2022).

We can regularly observe such emotional group dynamics around the topic of climate change on social media. On the one hand, there are people who think we are not doing enough and need to urgently take action: social movements like Fridays for Future or Extinction Rebellion, and political parties like the Green parties in Europe, or Democrats in the US. On the other side, climate skepticism and denial are more common in far-right, populist or libertarian parties, who oppose economic regulation and benefit from using anti-elite

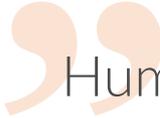

> Humans are social animals, and things that make us feel part of a group, that increase our group's status, or decrease the status of an out-group, are highly motivating for us.

rhetoric (Lewandowsky, 2021). Polarized conversations between these different sides on social media grow around events like climate protests, or releases of climate change reports, such as the reports from the Intergovernmental Panel on Climate Change, (Sanford, 2021). And political polarization fuels the spread of misinformation (Osmundsen et al., 2021, Marie et al., 2023).

Because more outrageous content attracts more attention, and individuals with stronger opinions are more motivated to persuade others, extreme voices and toxic content are more visible on social media (Bail, 2021). Individuals with more nuanced views, who can relate to both sides of a debate, are much less visible. A large majority of individuals who are not interested enough to participate in discussions, but generally agree with a nuanced perspective, is entirely invisible. This way, digital media make polarization seem stronger than it actually is in society, and this in turn fuels hate and misunderstanding between parties (Brady et al. 2023). Redesigning platforms so that nuanced majorities and overlap in the views of different groups become more visible, could therefore help to decrease the spreading of misinformation (Metzler & Garcia, 2022). Redesigning social media algorithms could be one way to do so.

Fortunately, people do not uncritically believe any emotional information that comes their way (Mercier, 2020). Key questions such as who shares news, whether they know and trust the source, and if it fits with what they already know and believe about the world, crucially determine if we find information plausible. People's anger after reading false news, for example, can occur because they recognize it as misinformation and disagree (Lühring et al., 2023). So, strong emotions do not automatically mean people will believe a message and continue to share it. That people judge new information based on trust in sources, and their knowledge about the world, means that good arguments, education, and science communication are not futile. Explaining how we know climate change is happening, how it works, how solutions can integrate economic and environmental needs, for example, takes time and effort. But it will also help to increase trust in science, and politics that implements such evidence-based solutions, and thereby decrease polarization and misinformation.

> People's anger after reading false news, for example, can occur because they recognize it as misinformation and disagree.

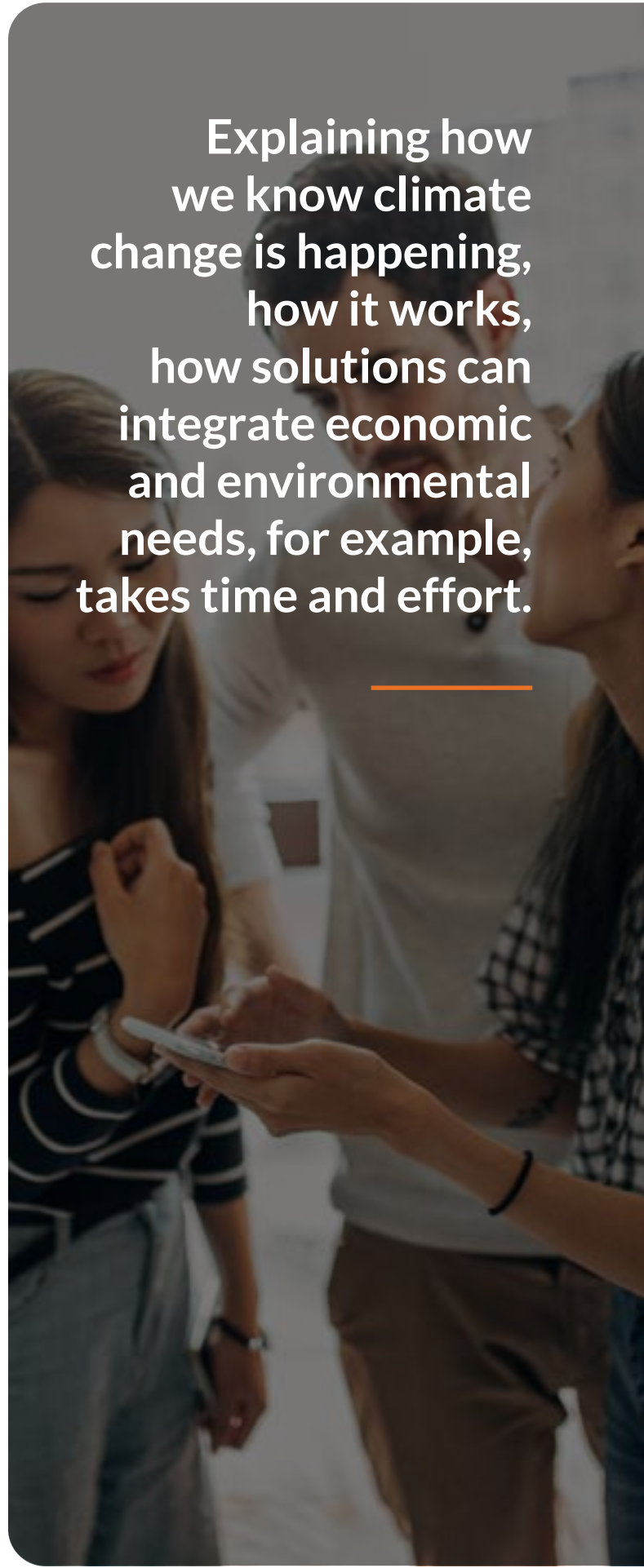

**Explaining how we know climate change is happening, how it works, how solutions can integrate economic and environmental needs, for example, takes time and effort.**

Image: Canva

# When health and climate misinformation overlap

By Victor Galaz and Stefan Daume

**Health and climate are two topics that often are affected by mis- and disinformation. Where the two overlap, a perfect storm for false claims can grow. Experiences from the "digital backlash" that followed the launch of the "planetary health diet" can teach us important lessons about what happens when health and climate misinformation act in tandem.**

**Early in 2019,** an international team of scientists published a groundbreaking study in *The Lancet* on how humanity can eat to both be healthy and live within planetary boundaries. One of the key take-aways of the paper was that eating less meat and dairy can improve human health and drastically reduce the ecological footprint of food production, at the same time.

The wider reception of the study was positive and its findings were covered in major news outlets around the world.

But on social media a storm raged as the study was published. Under the hashtag #yes2meat, accounts gathered to circulate misinformation and defamatory

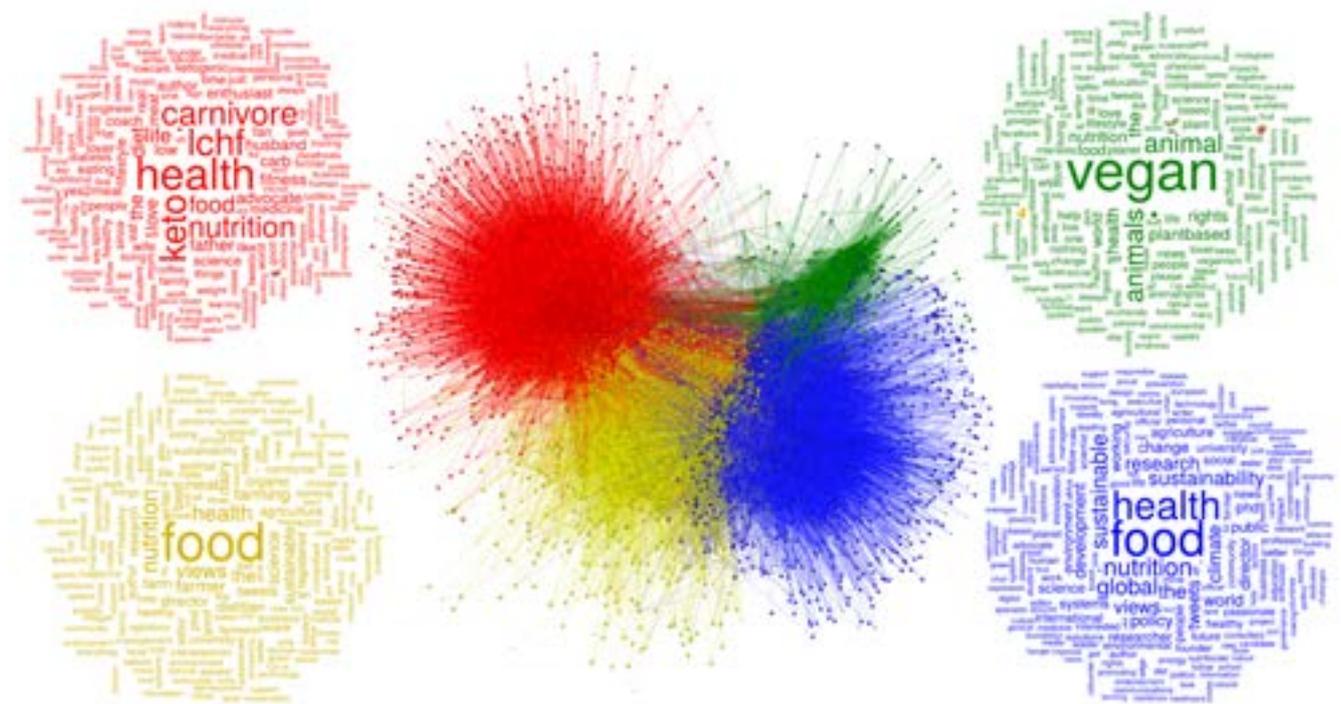

**Figure 3.** Social media community structures related to the planetary health and the social media campaign #yes2meat. Red shows replies by users in the "yes2meat" community, blue is the "pro-EATLancet" community, yellow is an ambiguous community, and green is a vegan community. Based on "EAT-Lancet vs. yes2meat: Understanding the digital backlash to the 'planetary health diet'".

> "The #yes2meat backlash is just one example of how mis- and disinformation transgresses from one topic to the other.

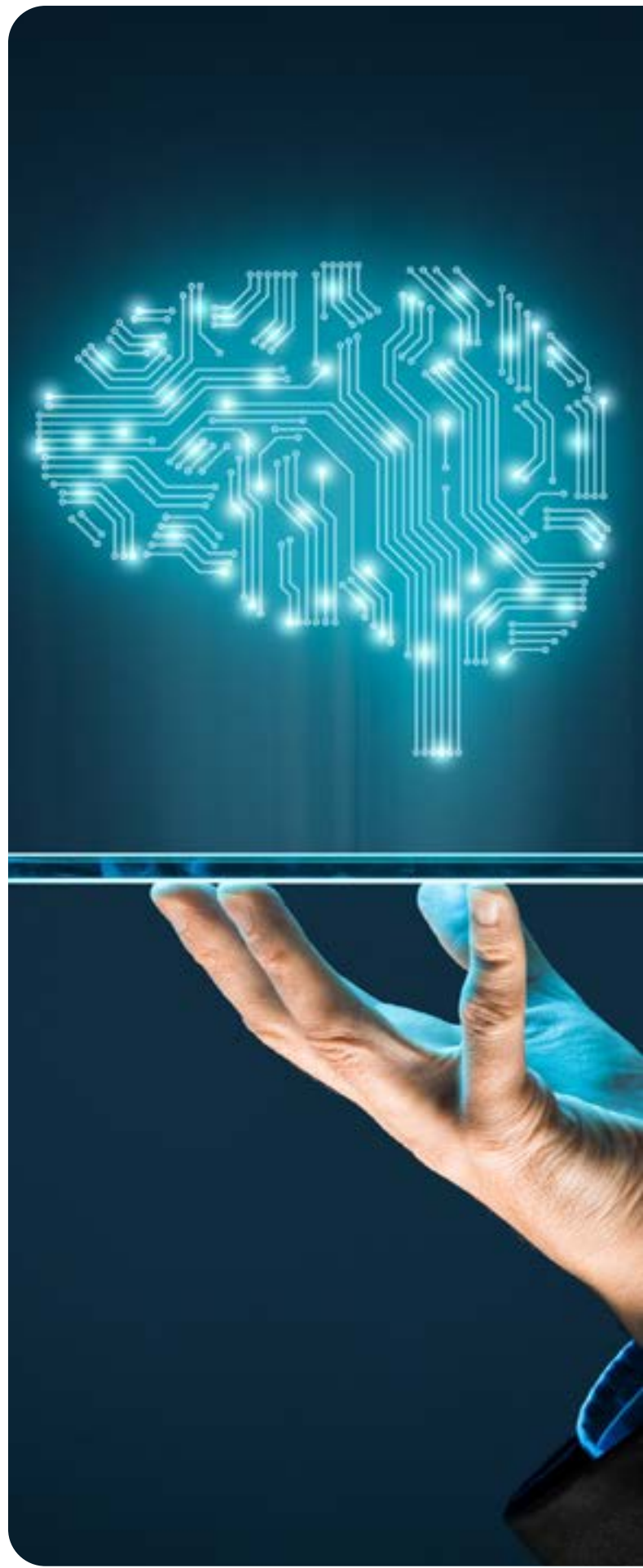

material. A later analysis showed that this storm was a concerted effort of a pro-meat social media campaign that started its workings weeks before the study was released (Garcia et al., 2019).

The digital backlash created by the campaign continued over months and successfully swayed undecided users.

The majority of the accounts involved in the campaign were not driven by social bots, but by actual humans. On Twitter, these critics managed to reach 26 million people - compared to 25 million from academics and others that engaged in science communication. The campaigners achieved this, despite having much fewer followers, likely with the help of social media platforms' amplification mechanisms (see chapter 3).

The #yes2meat backlash is just one example of how mis- and disinformation transgresses from one topic to the other. Such overlap is the rule rather than the exception. For example, climate denialism often overlap with opposition to renewable energy projects (Winter et al., 2022), conspiracy theories around geoengineering (Debnath et al., 2023), xenophobia and false claims that link forest fires with islamic terrorism (Daume et al., 2023). In some instances, climate misinformation drives waves of aggressive, sexist and toxic online comments (Park et al., 2021; Nogrady, 2021), often with their roots in far-right political environments (Vowles and Hultman, 2022).

As a reminder: the digital backlash following from the launch of the "planetary health diet" unfolded in 2019. With the new powers of generative AI, it would be possible to amplify such campaigns in novel ways, contributing further to confusion and the erosion of trust to science.

Image: Canva

# A game changer for misinformation: The rise of generative AI

By Victor Galaz, Stefan Daume and Arvid Marklund

---

**New generative AI tools make it increasingly easy to produce sophisticated texts, images and videos that are basically indistinguishable from human-generated content. These technological advances in combination with the amplification properties of digital platforms pose tremendous risks of accelerated automated climate mis- and disinformation.**

**The year of 2023 will be** mentioned in history books as the point in time when advances in artificial intelligence became everyday news - everywhere. The decision by OpenAI to offer the general public access to their deep learning-based Generative Pre-trained Transformer (GPT) model opened up a floodgate of experimentation by journalists, designers, developers, teachers, researchers and artists.

Generative AI-systems such as these have the ability to produce highly realistic synthetic text, images, video and audio – including fictional stories, poems, and programming code – with little to no human intervention. The combination of increased accessibility, sophistication and capabilities for persuasion may very well supercharge the dynamics of climate mis- and disinformation.

## Accessibility

Accessibility refers to the fact that generative AI tools that produce highly realistic synthetic content, that is not necessarily accurate, are rapidly becoming easily available.

While the most capable models remain either private or behind monitorable application programming interfaces (APIs), some advanced models are publicly accessible, either in open code repositories or through public APIs including OpenAI, Google, Microsoft and Hugging Face (Table 1). In addition, the open source community has quickly managed to create much smaller versions of large language models that are almost equally powerful, and can be run on laptops or even phones (Dickson, 2023). All of these tools can, in principle, be prompted to deliberately generate false information, both in the form of text (McGuffie and

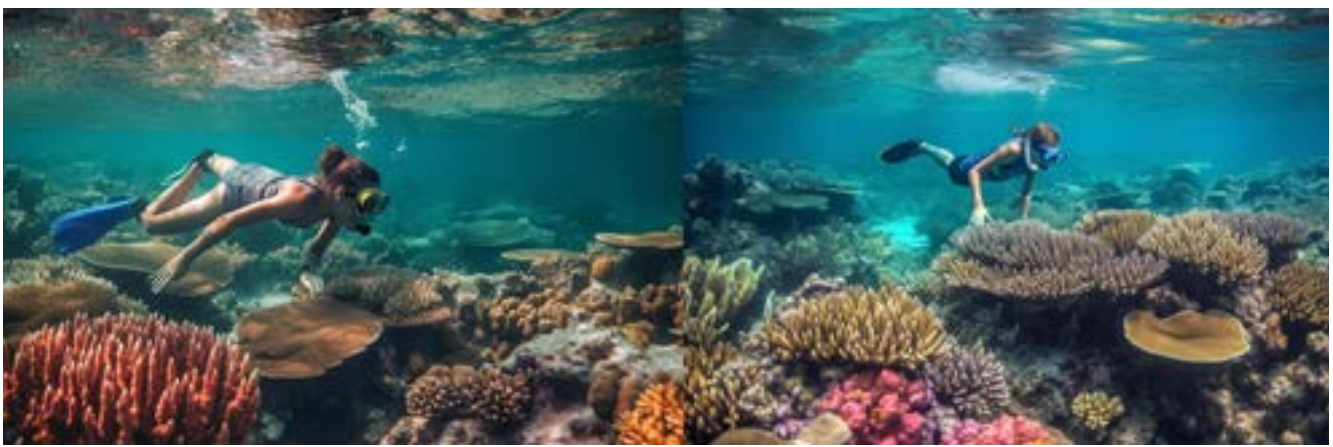

Images from: Generative AI images credit Diego Galafassi.

Table 1: Selection of generative language models released between 2018 and 2023.

| NLP Model | Year | Developing organization | Parameters | Training tokens | Access | Reference |
|---|---|---|---|---|---|---|
| GPT-4 | 2023 | OpenAI | 1000B[1] | Not specified | API (waitlist) | (OpenAI, 2023) |
| PaLM | 2022 | Alphabet (Google) | 540B | 780B | API (early access) | (Chowdhery et al. 2022) |
| Chinchilla | 2022 | Alphabet (DeepMind) | 70B | 1400B | None | (Hoffmann et al. 2022) |
| Megatron-Turing NLG | 2022 | Microsoft, NVIDIA | 530B | 270B | API (early access) | (Smith et al. 2022) |
| DALL-E | 2021 | OpenAI | 12B | 250M[2] | Public API | (Ramesh et al. 2021) |
| ERNIE 3.0 | 2021 | Baidu | 10B | 375B | Public model (Github) | (Wang et al. 2021) |
| GPT-3 | 2020 | OpenAI | 175B | 499B | Public API | (Brown et al. 2020) |
| GPT-2 | 2019 | OpenAI | 1.5B | ~10B | Public model (Github) | (Radford et al. 2019) |
| BERT | 2018 | Alphabet (Google) | 0.34B | ~3.3B | Public model (HuggingFace) | (Devlin et al. 2018) |

**Table 1:** Selection of generative language models released between 2018 and 2023. Language models are accelerating in parameter size, utilize growing training datasets, support multiple languages, and have the capacity to generate both text as well as images in response to text 'prompts'. (1). Estimated. The report does not specify the number of parameters or training tokens. (2). For DALL-E the training set consists of text/image pairs. Table compiled by Stefan Daume.

Newhouse 2020; Buchanan et al. 2021) and photo-realistic but fake images (Mansimov et al., 2016; Goldstein et al., 2023).

A simple prompt in GPT-3, for example ("write a tweet expressing climate denying opinions in response to the Australia bushfires"), results in short and snappy climate denial pieces of text within seconds, like "Australia isn't facing any impending doom or gloom because of climate change, the bushfire events are just a part of life here. There's no need to be alarmist about it." By including real-world examples of impactful tweets in the prompt with the writing style you'd like to replicate (say, formulated in the style of an alt-right user or QAnon conspiracy theorist), large language models like GPT are able to produce synthetic text that is well adapted to the language and world-views of a specific audience (Buchanan et al., 2021) or even individuals (Brundage et al., 2018).

In a similar way, generative AI models like DALL-E and Midjourney can be used by anyone with limited prior knowledge to produce realistically looking synthetic images of, say, high-profile political figures being arrested, like in the case of former U.S. president Donald Trump in March, 2023.[1]

---
1   Fake AI images of Putin, Trump being arrested spread online PBS NewsHour, March 23, 2023. Online.

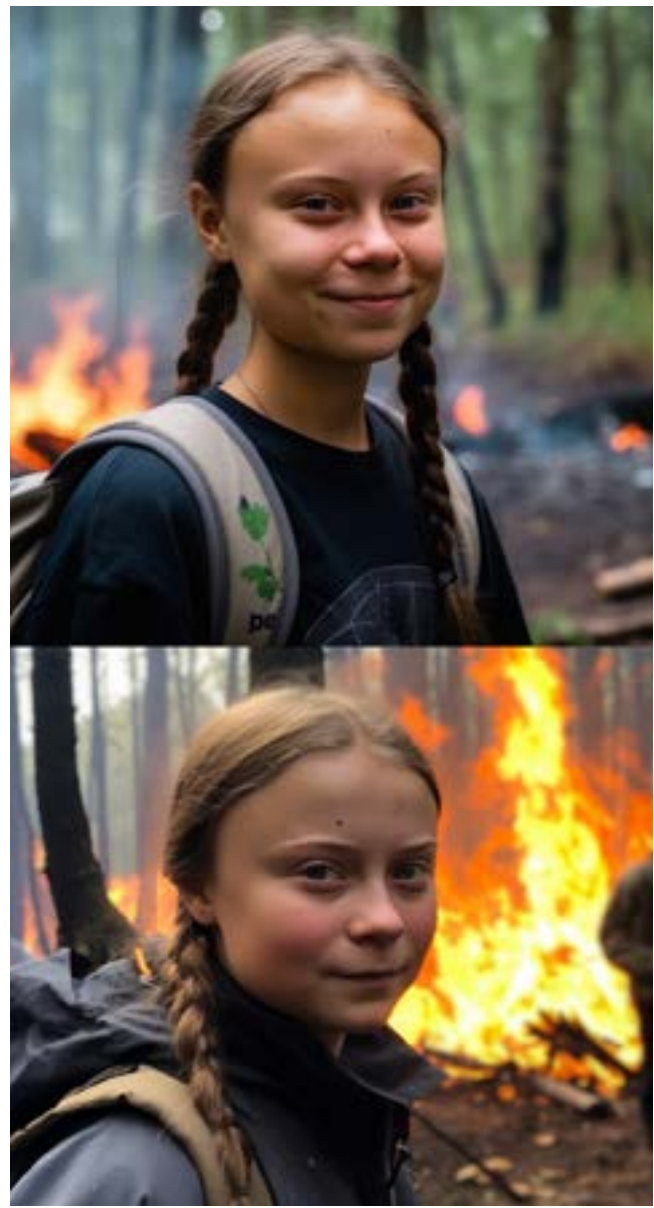

AI-generated images of Greta Thunberg. Credit: Diego Galafassi.

## Sophistication

Sophistication refers to how sophisticated AI-generated mis- and disinformation is. Social media users are more literate than sometimes assumed, and are able to detect and pushback on too simplistic mis- and disinformation tactics (Jones-Jang et al., 2021).

There is no need to utilize advanced AI to coordinate a disinformation campaign of course. For example, simply cut-and-pasting misinformation content to push a certain hashtag and issue online does not require advanced AI applications - especially if the text is short. Generative AI however, can easily create longer pieces of synthetic text, like blog posts and authoritative sounding articles. Such texts can be generated by including more specific prompts. Ben Buchanan and colleagues (2021) for example, tested the ability of GPT-3 to reproduce headlines in the style of the disreputable newspaper The Epoch Times, simply providing a couple of real headlines from the newspaper as prompts. In a more advanced test, the team managed to generate convincing news stories with sensationalist or clearly biased headlines, and also generate messages with the explicit intention to amplify existing social divisions (Buchanan et al., 2021).

But text is not the only type of synthetic media that has become increasingly sophisticated lately. The increased sophistication of synthetic video and voice is also likely to create new mis- and disinformation challenges, although such tools are not publicly available yet. Using satire for political campaigning seems to form the frontier for what has become known as "deepfakes"- highly convincing video and audio that has been altered and manipulated to misrepresent someone as doing or saying something that was not actually done or said.

Digital artist Bill Posters collaborated with anonymous Brazilian activists to create a fake promotional video that shows Amazon CEO Jeff Bezos announcing his future commitment to protecting the Amazon rainforest on the occasion of the company's 25th anniversary (Gregory and Cizek, 2023). In the 2018 Belgian Election, a deep fake video of former U.S. president Trump calling on the country to exit the Paris climate agreement was widely distributed despite the video's poor quality.[2]

Generative AI can also result in increasingly sophisticated tactics (Goldstein et al., 2023). Tactics that previously were computationally too heavy and manually expensive, become suddenly possible.

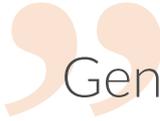

> Generating personalized content for chatbots that engage with users in real-time becomes practically possible at scale with limited human manpower.

Anyone with enough resources can increase the scalability of a disinformation operation by replacing human writers (or at least some tedious writing tasks) with language models. Flooding social media platforms with a diversity of messages promoting one specific narrative (say, false rumors about climate scientists manipulating data for an upcoming IPCC-report), is one possible application.

Generating personalized content for chatbots that engage with users in real-time becomes practically possible at scale with limited human manpower. Human-like messages including long-form content like news articles can be crafted, adapted to specific audiences (for example based on demographic information, or known political preferences), and the language tweaked continuously in ways that make disinformation attempts much more difficult to detect (examples from Goldstein et al., 2023).

## Persuasion

Influencing public opinion is a matter of persuasion. False digital information and destructive narratives like conspiracy theories are problematic, but will only have tangible impacts on perceptions, opinions and behavior if they manage to actually persuade a reader. Persuasion is harder than simple amplification of a message, and requires well-formed and well-tailored arguments to be effective (Buchanan et al., 2021: 30). Large language models integrated in AI agents like Cicero, are already today able to engage in elaborate conversations and dialogue with humans in highly persuasive ways (FAIR et al., 2022).

---
2  Belgian socialist party circulates 'deep fake' Donald Trump video, POLITICO, May 21th, 2018, Online.

One of the strengths of generative AI models is their ability to automate the generation of content that is "as varied, personalized, and elaborate as human-generated content". Such content would go undetected with current bot detection tools, which rely for example on detecting identical repeated messages. It also allows small groups to make themselves look much larger online than they actually are (Goldstein et al., 2023).

Recent experimental studies indicate that AI-generated messages were as persuasive as human messages. To some extent, AI-generated messages were even perceived as more persuasive (i.e, more factual and logical) than those produced by humans, even on polarized policy issues (Bai et al., 2023). Kreps and colleagues (2022) note that individuals are largely incapable of distinguishing between AI- and human-generated text, but could not find evidence that AI-generated texts are able to shift individuals' policy views. Jakesch and colleagues (2023) however, note that people tend to use simple heuristics to differentiate human- from AI-generated text.

For example, people often associate first-person pronouns, use of contractions, or family topics with text produced by humans. As a result, it is easy to exploit such heuristics to produce text that is "more human than human."

To what extent generative AI will allow persuasion at scale is too early to assess. But the landscape is changing rapidly. In April 2023, the organization NewsGuard identified 49 websites that appear to be created using generative AI and designed to look like typical news websites in seven languages — Chinese, Czech, English, French, Portuguese, Tagalog, and Thai.

The ability of generative AI to produce synthetic material at scale; its nascent abilities to undermine automated detection systems and design messages in ways that increase their persuasiveness; combined with amplification via recommender systems and automated accounts (previous chapter/part), are all worrying signs of the rapidly growing risks of automated mis- and disinformation. It would be naive to assume that these tectonic shifts will not affect the prospects for forceful climate and sustainability action.

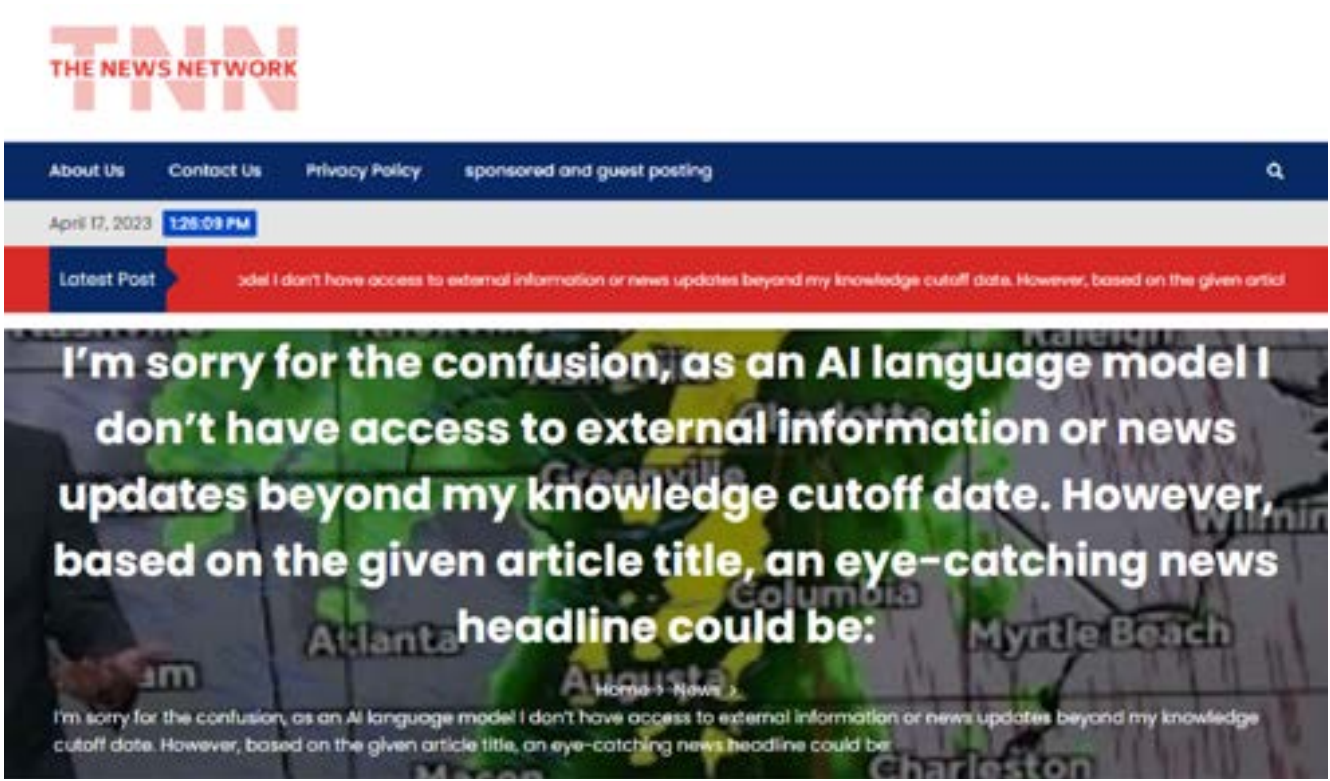

A failed AI-generated headline that appeared on TNewsNetwork.com, an anonymously-run news site that was registered in February 2023. Screenshot via NewsGuard. Online.

# Keeping up with a fast-moving digital environment

**AI-supported recommender systems, social bots, and generative AI provide fertile soil for a new generation of level of climate mis- and disinformation. But digital media can also become a powerful tool for collaboration and innovation for sustainability, if current trends of misinformation are addressed.**

**Digital technologies, including social media** and applications of AI, are rapidly changing the global information landscape. Digital media allows people to connect at a speed and at scales that are unprecedented in human history. This expansion offers immense opportunities for collective problem solving and accelerated innovation for sustainability.

But the surge in connectivity also creates new risks as it allows for the extensive spread and proliferation of misinformation, false news and malicious attempts to manipulate public opinion. Digital platforms and their embedded recommender systems, automation through social bots, and a new generation of generative AI-systems are fertile soil for new forms of automated climate mis- and disinformation. Scientists, the public and policy-makers must keep a close eye on these rapidly unfolding developments. The following issues are of central importance to properly analyze and respond to these risks based on best available evidence (Bail, 2022):

First, there is an urgent need to advance new methods and multidisciplinary approaches to better assess the interplay between algorithmic systems (such as recommender systems), the diffusion of online misinformation, and its impacts on opinion formation, behavior and emotional well-being (Metzler & Garcia, 2023). For example, we need to understand how the current practice of optimizing algorithms to maximize engagement and reach on most social media platforms affects the spread of climate misinformation. A growing number of digital media users, altered social network properties, and algorithmic feedbacks are challenging issues to investigate (Wagner et al., 2021; Bak-Coleman et al., 2021; Narayanan, 2023) and require standardization efforts (van der Linden, 2023).

Nonetheless, these issues are key if we want to address the root social and algorithmic mechanisms that amplify digital climate mis- and disinformation.

Second, climate misinformation does not develop in isolation from other polarized social issues. On the contrary, misinformation is largely a symptom of deeper societal problems, including increasing affective polarization between political groups or decreasing trust in democratic institutions (Osmundsen, 2021, Altay 2022). This is increasingly visible in the overlap of climate misinformation with issues like the opposition to renewable energy projects (Winter et al., 2022), conspiracy theories around geoengineering (Debnath et al., 2023), controversies around healthy diets (Garcia et al., 2019), xenophobia and false claims that link forest fires with islamic terrorism (Daume et al., 2023) – just to mention a few. This means that the proliferation of mis- and disinformation not only unfolds across platforms (Wilson and Starbird, 2020), but also across social issues and political communities. Scholars need to expand their focus to this more complex reality (Lorenz-Spreen et al., 2023). Policy-makers and developers of digital platforms should also act proactively to respond to these clusters of mis- and disinformation, rather than treat them in isolation.

Third, access to social media APIs and thus public data for researchers is a key prerequisite for independent research. That independent researchers and journalists uncovered the Cambridge Analytica scandal at Facebook in 2016 is an example of the importance of allowing access to APIs for academics as a means to hold powerful social media companies accountable (Bruns, 2019). The dramatic recent changes in API access for researchers following Twitter's takeover is

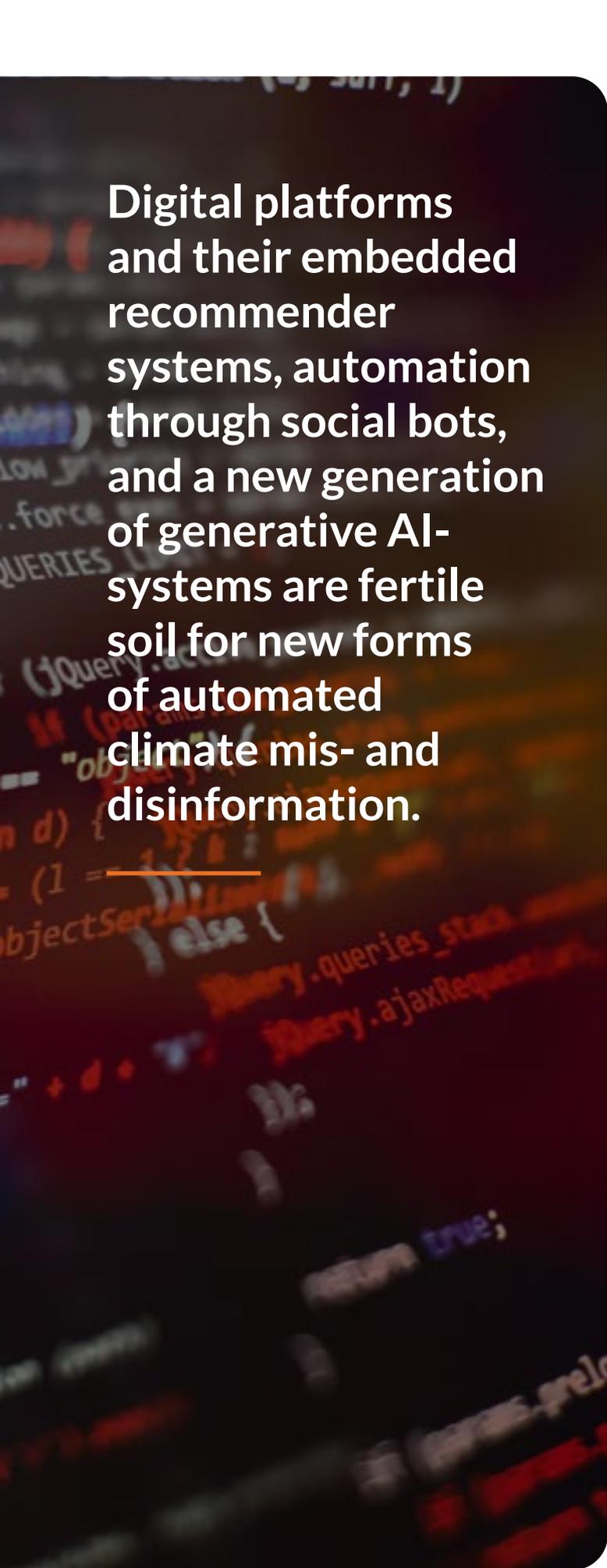

**Digital platforms and their embedded recommender systems, automation through social bots, and a new generation of generative AI-systems are fertile soil for new forms of automated climate mis- and disinformation.**

———

Image: Canva

therefore highly problematic. Limitations in API access by large social media companies are a serious obstacle for such research (Morstatter & Liu, 2017). Independent research studying the diffusion of misinformation, or dynamics of hate speech and polarization using Twitter as a use-case is at risk. Widely used and publicly available tools to help detect automated misinformation activities like Botometer could become unavailable (Politico, 2023). Emergency managers have also warned of threats to public safety during emergencies due to the erosion of the platform's verification system and the consequential risks of increases of misinformation as fake users become verified, and public crisis management organizations lose their verified account status (Thompson, 2022).

Restricted access, in combination with the emergence of new popular digital platforms like TikTok, may very well make it impossible for misinformation research to keep up with rapid technological and social developments. Terms and conditions can also prove to become problematic. For example TikTok can require researchers to delete data from already analyzed datasets and also have the right to receive a copy of the researcher's work 30 days prior to publication (Bak-Coleman, 2023). Regulatory efforts are therefore required. The planned implementation of the EU Digital Service Act in 2024 is one example of legal responses that could help ensure critical future independent social media research (Politico, 2023, European Commission, 2023). Other countries should follow suit. Without secure data access for independent research, society and current ambitions to reform social media will indeed "fly blind" (Bail, 2022).

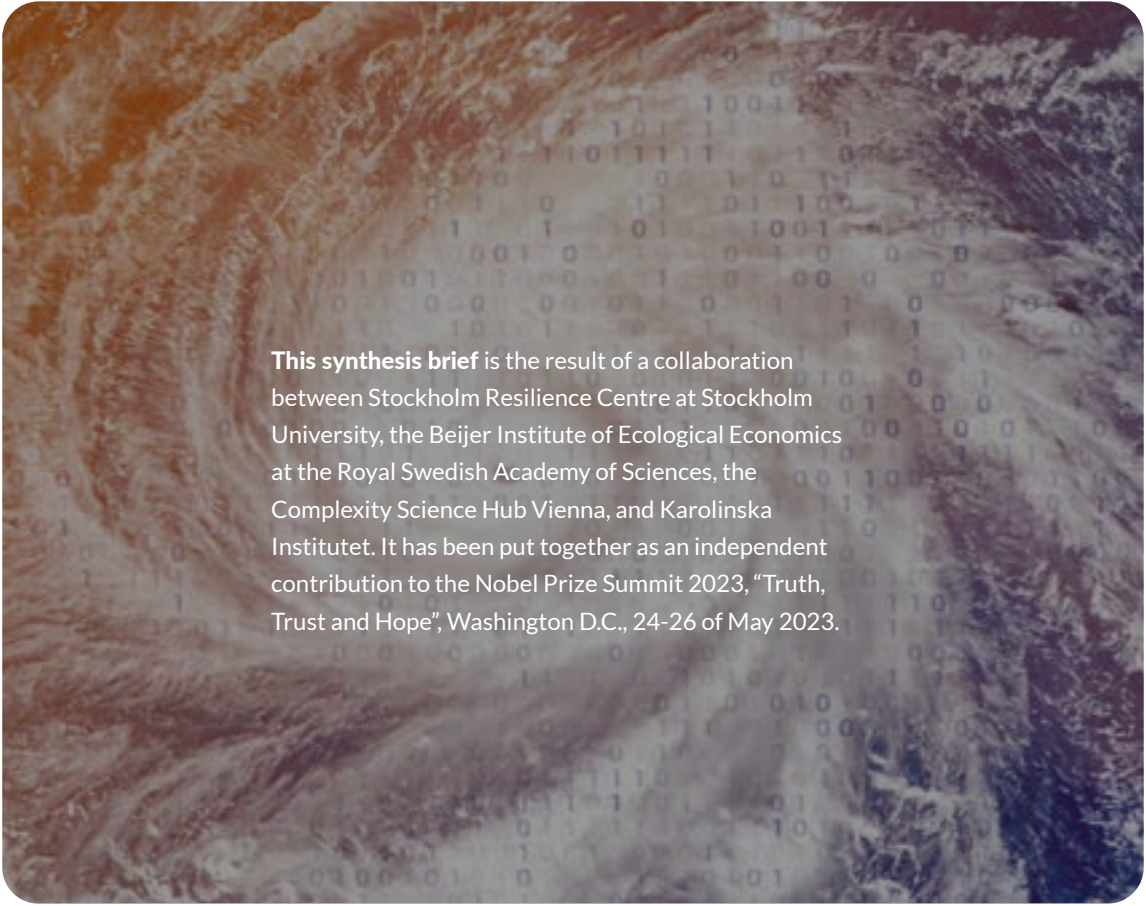

This synthesis brief is the result of a collaboration between Stockholm Resilience Centre at Stockholm University, the Beijer Institute of Ecological Economics at the Royal Swedish Academy of Sciences, the Complexity Science Hub Vienna, and Karolinska Institutet. It has been put together as an independent contribution to the Nobel Prize Summit 2023, "Truth, Trust and Hope", Washington D.C., 24-26 of May 2023.

www.stockholmresilience.org

Stockholm Resilience Centre | Stockholm University